\documentclass[reprint,superscriptaddress,a4paper]{revtex4-2}
\usepackage[utf8]{inputenc}

\usepackage{mathptmx,graphicx}

\usepackage{soul}
\usepackage{wrapfig}
\usepackage{amsmath}
\usepackage[export]{adjustbox}
\usepackage{float}
\usepackage{chemformula}

\usepackage[colorlinks=true, allcolors=blue]{hyperref}
\usepackage[T1]{fontenc}
\usepackage{textgreek}
\usepackage{csquotes}
\usepackage{comment} 

\usepackage{wasysym} % for the diameter command 
\usepackage{mathtools} % make absolute value symbol

\begin{document}

\title{Surface lattice resonance lasers with epitaxial InP gain medium}

\author{Anna Fischer$^*$}
\affiliation{Blackett Laboratory, Department of Physics, Imperial College London, London, UK}
\affiliation{IBM Research Europe - Z\"{u}rich, S\"{a}umerstrasse 4, R\"{u}schlikon, 8803, Switzerland}
\author{Toby Severs Millard$^*$}
\affiliation{Blackett Laboratory, Department of Physics, Imperial College London, London, UK}
\affiliation{National Physical Laboratory, Teddington TW11 0LW, United Kingdom}
\author{Xiaofei Xiao}
\affiliation{Blackett Laboratory, Department of Physics, Imperial College London, London, UK}
\author{T.V. Raziman}
\affiliation{Blackett Laboratory, Department of Physics, Imperial College London, London, UK}
\affiliation{Department of Mathematics, Imperial College London, London, UK}
\author{Jakub Dranczewski}
\affiliation{Blackett Laboratory, Department of Physics, Imperial College London, London, UK}
\affiliation{IBM Research Europe - Z\"{u}rich, S\"{a}umerstrasse 4, R\"{u}schlikon, 8803, Switzerland}
\author{Ross C. Schofield}
\affiliation{Blackett Laboratory, Department of Physics, Imperial College London, London, UK}
\author{Heinz Schmid}
\affiliation{IBM Research Europe - Z\"{u}rich, S\"{a}umerstrasse 4, R\"{u}schlikon, 8803, Switzerland}
\author{Kirsten Moselund}
\affiliation{Paul Scherrer Institut, Forschungsstrasse 111, Villigen, 5232, Switzerland}
\affiliation{EPFL, Lausanne, 1015, Switzerland}
\author{Riccardo Sapienza}
\affiliation{Blackett Laboratory, Department of Physics, Imperial College London, London, UK}
\author{Rupert F. Oulton}
\affiliation{Blackett Laboratory, Department of Physics, Imperial College London, London, UK}

\begin{abstract}
$^*$ These authors contributed equally to the work. \\

Surface lattice resonance (SLR) lasers, where gain is supplied by a thin film active material and the feedback comes from multiple scattering by plasmonic nanoparticles, have shown both low threshold lasing and tunability of the angular and spectral emission. However, typically used materials such as organic dyes and QD films suffer from photo-degradation which hampers practical applications. Here, we demonstrate photo-stable single-mode lasing of SLR modes sustained in an epitaxial solid-state InP slab waveguide.  
The nanoparticle array is weakly coupled to the optical modes, which decreases the scattering losses and hence the experimental lasing threshold is as low as 90 $\mu$J/cm$^{2}$. The nanoparticle periodicity defines the lasing wavelength and enables tuneable emission wavelengths over a 70 nm spectral range.
Combining plasmonic nanoparticles with an epitaxial solid-state gain medium paves the way for large-area on-chip integrated SLR lasers for applications including optical communication, optical computing, sensing, and LiDAR.

\end{abstract}

\maketitle

\subsection*{Introduction }

On-chip lasers in combination with integrated photonic circuits are promising for the next generation of optical communication systems~\cite{wang_2017, li_2022}, chemical and biological sensors~\cite{Altug_Oh_Maier_Homola_2022}, optical computation~\cite{feldmann_2021, skalli_2022, Sludds_Bandyopadhyay_Chen_Zhong_Cochrane_Bernstein_Bunandar_Dixon_Hamilton_Streshinsky_2022}, and LiDAR systems~\cite{zhou_2023}. While Si based photonic integrated circuits benefit from low-cost, large-scale production while offering high speeds, band-width and efficiency, they are limited by the integration of suitable gain media~\cite{li_2022,yang_2023}.

Surface lattice resonance (SLR) lasers are of rising interest, as they 
show large-area, directional, and controllable mode emission~\cite{Kravets2018_review}, as well as broad post-fabrication control in the form of optical~\cite{Taskinen2020_SLR_opticalcontrol} and magnetic~\cite{Freire-Fernández2022_SLR_magSwitch} switching, mechanical mode tuning~\cite{Gupta_Probst_Goßler_Steiner_Schubert_Brasse_Konig_Fery_2019} and thermal mode control~\cite{Kelavuori_Vanyukov_Stolt_Karvinen_Rekola_Hakala_Huttunen_2022}. 
SLRs are formed by the hybridisation of the collective mode of a nanoparticle array and a diffractive surface wave, forming a cavity \cite{Kravets2018_review}. Similar to coupled resonator optical waveguides \cite{Wood2015_CROWs}, they can be engineered to achieve high resonant quality factors. In-plane SLR arrays, where nanoparticles are polarised in the sample plane, have been carefully developed to reach quality factors above $2\times 10^3$ that rival those of bound-states in the continuum (BICs)~\cite{Bin-Alam2021_ultraHighQ}.

Most SLR lasers have so far relied on organic dye molecules as gain media which are difficult to integrate with solid-state photonic integrated circuits~\cite{Schokker_Koenderink_2014,Wang2018_SLR_tunable_stretchy, Pourjamal2019_SLR_Ni_dyeMole}. 
Alternatively to dyes, semiconductor quantum dots (QDs) have been successfully used to achieve low threshold lasing~\cite{Guan2020_QD_SLR,Guan2020_SLR_QD_pol,Winkler2020_SLR_QD_DualWL}. 
However, QDs~\cite{Guan2020_QD_SLR,Guan2020_SLR_QD_pol, Winkler2020_SLR_QD_DualWL} are subject to photo-degradation, being more susceptible to oxidation due to their large surface area. This is a well known issue for nanolaser devices, with organic dye molecules~\cite{Schokker_Koenderink_2014,Wang2018_SLR_tunable_stretchy, Pourjamal2019_SLR_Ni_dyeMole} and perovskites~\cite{Huang2021_SLR_Perov,Xing2022_SLR_PerovQD} degrading similarly \cite{Vellaichamy2023_GM_photobleach}. 
A solid-state SLR laser with lanthanide-based upconverting nanoparticle gain medium has been demonstrated with continuous-wave pumping, although, the nanoparticle film limits integration with Si platforms \cite{Fernandez-Bravo_2019_CW}.

Here, we introduce the first SLR laser with solid-state epitaxial semiconductor gain medium that can be incorporated defect-free on-chip through wafer-bonding of InP to \ch{SiO2} on Si~\cite{caimiHeterogeneousIntegrationIII2021}. 
We place the nanoparticle outside of the gain medium waveguide forming a plasmonic-photonic hybrid SLR mode \cite{Wood2015_CROWs}.
The cylindrical plasmonic resonators are suspended in \ch{SiO2} and coupled to a slab waveguide of \ch{SiO2}/\ch{InP}/\ch{SiO2}. This has the benefit of optically accessing the InP gain medium without complicated fabrication, and 
reduces plasmonic absorption losses. We demonstrate low threshold, single-mode emission and spectral control of devices across large lateral areas, and further show that InP is robust to photo-degradation under prolonged pumping.

\subsection*{Results} 

\begin{figure*}[ht!]
    \centering
    \includegraphics[width=\linewidth]{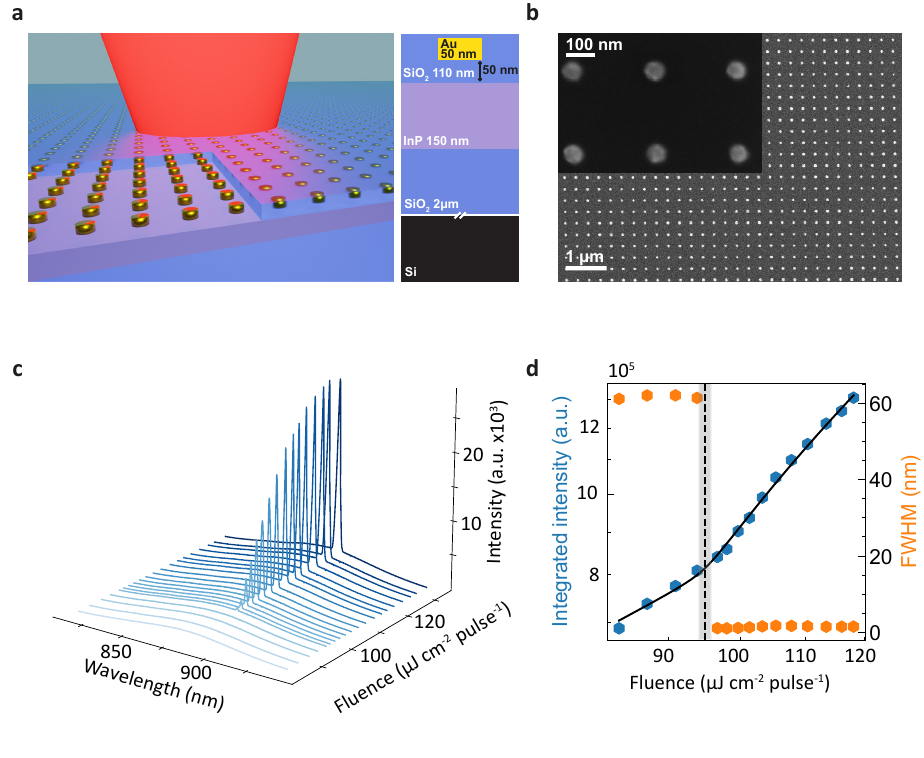}
    \caption{\textbf{Single-mode lasing from the Au nanoparticle array on InP}
    (a) Graphical illustration of the sample showing the \ch{SiO2} substrate (dark purple), InP layer (150 nm, light purple), \ch{SiO2} layer (110 nm, transparent dark purple) with embeded Au nanoparticle array (gold). 
    (b) SEM image of the Au nanoparticle array on InP with nanoparticle diameters of 60 nm and a period of 308 nm. The inset shows the same device at a higher magnification.
    (c) The photoluminescence and lasing spectra from the array in (b) at different pump fluences, indicating single-mode lasing. 
    (d) Integrated intensity versus pump fluence (LL-curve) in logarithmic scale (left axis, blue) showing a characteristic S-shape with a rate equation based fit (black line). The right axis shows the narrowing of the FWHM upon lasing (orange).}
    \label{fig:1}
\end{figure*}

The SLR laser is composed of a slab waveguide of 150 nm thick InP on a \ch{Si}/\ch{SiO2} substrate with a square 100x100 $\mu$m array of cylindrical Au nanoparticles, suspended in the second layer of 110 nm thick \ch{SiO2}, 50 nm above the InP. 
Figure \ref{fig:1}a shows a graphical representation of a typical SLR laser device measured, with a 2D cross-section shown in the inset. An SEM image of a nanoparticle array with 60 nm diameter and a period of 310 nm is shown in Figure~\ref{fig:1}b. 

Normally incident laser light at 633 nm excites the InP significantly above the 918 nm room temperature bandgap \cite{InP_bandgap_Pavesi_1991} and photoluminescence (PL) is subsequently emitted. A portion of the PL that is emitted in the sample plane is contained within a waveguide mode through coupling to the SLR mode created by the nanoparticle array. In turn, the gain medium experiences population inversion, for the wavelength of light trapped in said mode, and begins to lase. The 50 nm \ch{SiO2} spacing layer between the nanoparticles and InP acts to control the degree of coupling, allowing for optimisation of the laser process. Laser light is generated in the SLR mode, which couples to the far field via scattering from the nanoparticle array. From this qualitative description of the device mechanism, it is intuitive that the properties of the lasing mode can be tuned via appropriate tailoring of the InP waveguide gap, the \ch{SiO2} spacing layer and SLR array dimensions. A detailed description of the fabrication can be found in the \textit{Methods} section.

\begin{figure*}[ht!]
    \centering
        \includegraphics[width=\linewidth]{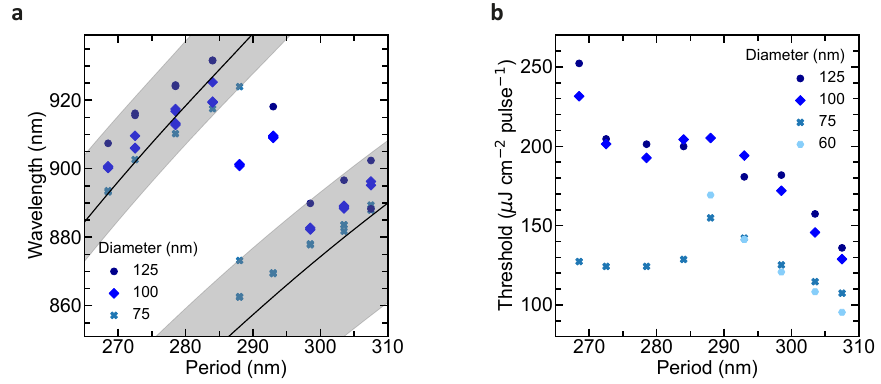}
    \caption{\textbf{Array period defines lasing peak wavelength.}
    (a) Peak wavelength versus Au nanoparticle array period for different nanoparticle diameters shows how the lasing mode red-shifts for increased periods. The plotted wavelengths are the peak wavelengths of devices just above threshold. The black lines are obtained from analytical theory with an InP thickness of 165 nm and a carrier-induced refractive increase of 10\%. The gray shaded area gives a range of possible analytical values for InP thicknesses between 150-165 nm, and refractive index changes of 8-13 \%.
    (b) Lasing threshold versus period for arrays of nanoparticles with different diameters, showing lowest threshold for 75 and 60 nm diameter particles.
    }
    \label{fig:2}
\end{figure*}

On exciting the device from figure \ref{fig:1}b with increased fluence, the emission spectra develop a sharp single peak, as shown in figure \ref{fig:1}c. Simultaneously, the broad PL emission increases linearly with fluence and red shifts due to localised Varshni heating \cite{InP_BG_TempDep_Varshni1967} of InP. The transition to lasing is characterised by the appearance of the single sharp peak within the PL band, with an amplitude that increases non-linearly with pump fluence. The threshold of this lasing peak is clearly shown in the logarithmic LL curve of figure \ref{fig:1}d; highlighted by a higher order dependence of the integrated intensity (blue) on fluence and a clear drop in the FWHM (orange) of the highest intensity peak. The grey dashed line and box mark the lasing threshold of 94.99 $\pm$ 0.82 uJ cm$^{-2}$ pulse$^{-1}$---the lowest threshold seen across the devices---extracted by fitting rate equations that account for both spontaneous and stimulated emission \cite{Schofield_2023}. It is important to note here that the intensity has been integrated over a 60 nm window around the spectral position of the laser peak; a necessary filter to observe lasing against the broad PL emission. 
The observed threshold range is comparable to the lowest reported SLR laser threshold of 19 $\mu$J cm$^{-2}$ pulse$^{-1}$ coming from perovskite QDs~\cite{Xing2022_SLR_PerovQD}. Other gain media with comparable arrays show higher thresholds, with organic dyes ranging between 300 - 1500 $\mu$J cm$^{-2}$ pulse$^{-1}$ \cite{Taskinen2020_SLR_opticalcontrol, Schokker_Koenderink_2014}, core-shell QDs between 160 - 1000 \cite{Winkler2020_SLR_QD_DualWL, Guan2020_QD_SLR}, and lanthanide-based nanoparticles show thresholds of 400 $\mu$J cm$^{-2}$ pulse$^{-1}$ under pulsed-excitation~\cite{Fernandez-Bravo_2019_CW}.

We estimate the quality factor from the mode linewidth of this device to be around 800 comparing well with other plasmonic in-plane SLR cavities \cite{Bin-Alam2021_ultraHighQ}.
This may under-represent the actual performance of the cavity, as wavelength chirping from the optical modulation of the semiconductor carrier density introduces artificial broadening. Additionally, the PL emission of InP exhibits a blue shift due to a significant increase in carrier density under high pump fluence, further contributing to broadening \cite{Bennett1990_RIchange_carrierDensity_InP,Tiwari2021_metalCladInPNanodisks, Tiwari2022_InPMicroDisks_AuAnt}.
Despite the PL shift, the SLR keeps the lasing peak position well-anchored with only a <2 nm blue shift; comparable to the 1.1 nm FWHM. 

\begin{figure*}[ht!]
    \centering
    \includegraphics[width=\linewidth]{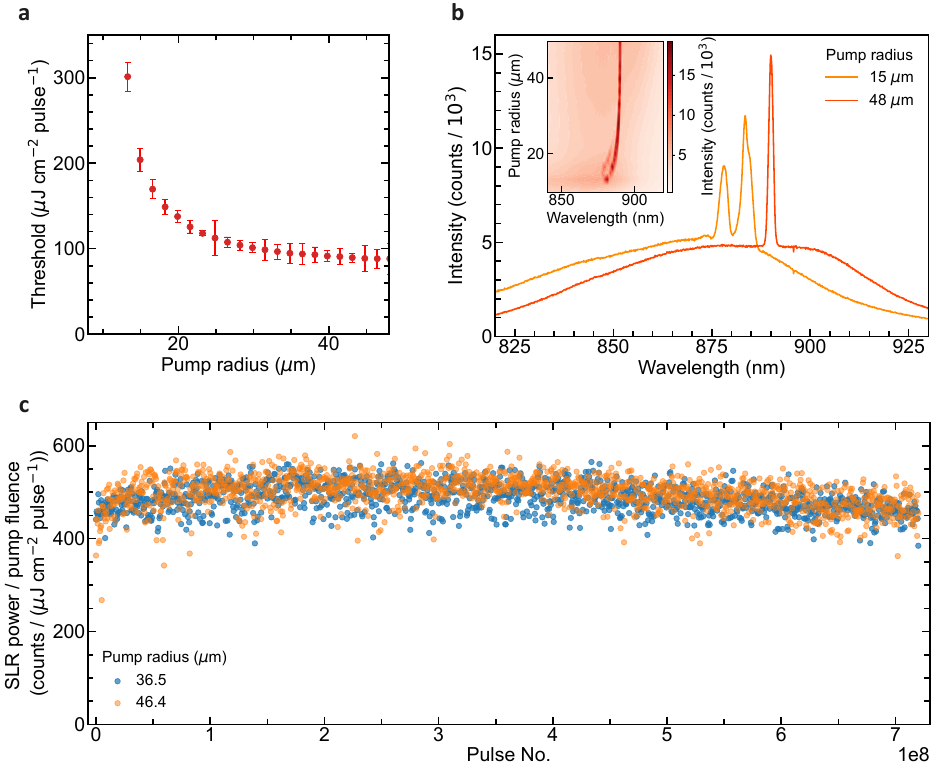}
    \caption{\textbf{Threshold and mode dependence on pump beam radius and photo-stability of InP gain medium to prolonged pumping.  }
    (a) Lasing threshold versus pump radius for a device with 60 nm diameter nanoparticles and a period of 308 nm. The threshold converges to a minimum value of 94.69 $\pm$ 2.22 $\mu$J cm$^{-2}$ pulse$^{-1}$ with increased pump radius. (b) Above-threshold spectra for the same device as in (a) showing single mode emission for pump radii larger than the characteristic cavity size and multimode emission below.
    (c) Prolonged pumping over two hours (7x10${^8}$ pulses) showing lasing intensity to be photo-stable within 6.1 \% (7.1 \%) for pump diameter of 36.5 \textmu{}m (46.4 \textmu{}m). }
    \label{fig:3}
\end{figure*}

Here, the hybrid SLR and slab waveguide mode defines the lasing wavelength, where it is dependent on the array period, nanoparticle diameter, and the InP thickness. 

The broad PL region of InP allows the lasing mode to be tuned over a spectral range of \(\sim\)70 nm, with the laser emission red-shifting with increasing diameter and period; as shown by the peak wavelengths just above threshold in figure~\ref{fig:2}a. As the period increases to ~290 nm the SLR mode red-shifts beyond the InP PL region and another mode appears to lase at shorter wavelengths. 
We can distinguish three modes across our devices; two more prominent modes and a third mode in-between these (Figure~\ref{fig:2}a). 
This middle mode is most prominent at periods of \(\sim\)290 nm. For these periods, the wavelengths of the two main modes are at either edge of the InP gain region, which is centered around 900 nm. As the main modes experience less gain, the lossier third mode can reach threshold due to lower gain competition. At higher pump powers, the middle mode also reaches threshold for other periods (SI figure 1).

To theoretically describe the two main modes, we analytically calculate the grating condition for coupling to the TE and TM modes of the system assuming normal emission of the modes, as is typical for SLR lasers with weak coupling \cite{Schokker_Koenderink_2014} (see Supporting Information SII).This analytical method was validated by comparison with finite element method (FEM) simulations. To obtain agreement between the theoretical prediction and the experimental observation of the mode wavelengths, we tuned the values for the thickness and the refractive index of the InP waveguide around the expected values. Increasing both properties by 10\% gives a good overlap with the experimental results (black lines in figure~\ref{fig:2}a). A 10\% mismatch in waveguide thickness can likely occur during the MOCVD growth of the InP layer. The increase in refractive index can be explained due to band filling and plasma dispersion effects typical for semiconductors at high carrier densities \cite{Bennett1990_RIchange_carrierDensity_InP}. These carrier-induced changes in refractive index are reinforced by the very short pump pulses and hence high peak powers used in our experiments.

We use finite-difference time-domain simulations to consider the bandstructure of a typical device with increasing $k_\parallel$ and observe an intersection of the TE and TM modes that occurs at a higher angle (see Supporting Information SIII).
This crossing leads to the formation of a third mode with expected oblique emission. 
Tuneable directional emission has previously been enabled through lattice engineering, which moves high-symmetry point modes into the gain region of the active material \cite{Guan2020_QD_SLR}. In our work, the presence of distinct TE and TM modes due to the hybrid SLR and waveguide mode leads to additional crossings in-between symmetry points and hence closer spaced cavity modes.

The scattering and absorption by the plasmonic nanoparticles are strongly dependent on their size, leading to lower lasing thresholds for devices with 60 and 75 nm diameter cylindrical nanoparticles, compared to 100 and 125 nm.
The reflectance peaks of nanoparticles with diameters of 100 and 125 nm are centred at $\sim$600 and $\sim$640 nm, respectively. The spectral overlap with the 633 nm pump laser reduces the amount of light reaching the InP, increasing the lasing threshold. Given the broadband nature of plasmonic resonances, there is also 
increased absorption at the the operating wavelengths of the devices. This reduces the quality factor of the cavity and increases the lasing threshold. 
Simultaneously, larger particles scatter light more strongly, allowing light to escape the cavity more easily. 
While the devices with 75 nm and 60 nm nanoparticles show similar low thresholds for the TE SLR mode, the 60 nm nanoparticle devices show no lasing in the TM SLR mode for periods below 290 nm. 
In this range, resonant modes and amplified spontaneous emission (ASE) were observed, but the devices did not reach a laser threshold. 
This is due to the in-plane polarising field of the TE mode coupling better to the nanoparticles than the TM mode. 
One should note that the threshold cannot be reduced indefinitely by reducing particle size. Minimising absorption but maintaining sufficient scattering are both important.
Besides the threshold dependence on nanoparticle diameter, there is also a dependence on period. For devices with periods around 290 nm, the thresholds show a small increase. This increase is likely due to gain competition between the modes as periods in this region support multimode operation (Figure~\ref{fig:2}a,b). For periods $>$295 nm the threshold decreases due to reduced absorption in InP as the lasing mode red-shifts \cite{Bennett1990_RIchange_carrierDensity_InP}.

Theoretically, SLR laser arrays are considered to be infinite in size in order to simplify the system analysis. However, experiments are limited by the pump beam radius used to excite the array. By increasing the beam radius the characteristic properties of the laser converge to values that would be achieved in an infinite system. Figure \ref{fig:3} a and b demonstrate this convergence for lasing threshold and number of lasing modes, respectively.
Within the errors of the experimental data, the device can be said to have converged to a threshold value of 94.69 $\pm$ 2.22 $\mu$J cm$^{-2}$ pulse$^{-1}$ at radii $\geq$ 29.85 $\mu$m, thereby defining the minimum effective cavity diameter of the laser. 
For this estimation the larger error bar on the 24.88 $\mu$m measurement has been ignored as an outlier. This effective cavity approximation is a measure of the coherence length of the particle-waveguide interaction of the SLR modes and shows a weak interaction over \(\sim\)100 nanoparticles. 

Spectra highlighting the development of the lasing mode with increased pump radius, figure \ref{fig:3}b, shows reasonable agreement with our estimate for the effective cavity size. At smaller pump radii, where lasing is seen, there is typically a second and third peak competing for gain in the system. As the pump radius increases from 20 $\mu$m these extra peaks diminish and the system becomes single mode from 26.54 $\mu$m. Further increasing the pump spot radius results in no change to the spectral shape of emission. For smaller pump spot radii, the inset of figure \ref{fig:3}b shows an observable blue shift and the emergence of multiple modes. These features, alongside increased spacing between these multiple modes with decreasing pump spot radius, are consistent with the presence of lateral confinement linked to pump spot size \cite{Wood2015_CROWs}.
The origin of the lateral confinement is likely to be carrier density-induced nonlinear refraction, where the centre of the pump beam has a higher refractive index than surrounding un-pumped regions. The potential created by the refractive index gradient leads to the quantisation of the cavity and multiple equally spaced modes

A critical limitation of previously demonstrated SLR lasers is their susceptibility to photo-bleaching under prolonged operation. A recent study on the photo-stability of liquid crystal gain media characterised multiple materials including those commonly used for SLR lasers, such as organic fluorescent dyes, perovskites, CdSe/ZnS quantum dots and CdSe/CdS quantum rods \cite{Vellaichamy2023_GM_photobleach}. Of these, the largest photo-stable half-life observed was 3.8x10$^9$ pulses at 0.403 GW/cm$^2$ peak power density for CdSe/CdS quantum rods. Epitaxial bulk semiconductors are not expected to suffer the same degradation, as is indicated by the photo-stability plot in figure \ref{fig:3} (c). Here, a 200 fs laser was used to deliver 7x10$^8$ pulses over two hours to the 60 nm diameter 308 period device. The measurement was taken twice at 36.5 $\mu$m and 46.4 $\mu$m pump laser radii and at 0.516 and 0.472 GW/cm$^2$ peak power densities, respectively. The SLR lasing intensity is normalised to the pump power density at each measurement to remove the effect of input laser power fluctuations. The resulting two data sets show no signs of reduced performance and have only a 6.2\% (36.5 $\mu$m) and 7.1\% (46.4 $\mu$m) standard deviation from their respective means. In comparison, the emission intensity of the CdSe/CdS device mentioned above, has already dropped by $~ 40 \%$ for this number of pulses. This highlights the extremely stable lasing performance for our InP waveguide-coupled SLR laser.

\subsection*{Conclusion} 

In conclusion, we have demonstrated an SLR laser with an epitaxial solid-state active material of InP and a novel design that delocalises the cavity from plasmonic nanoparticles.
This device structure is theoretically optimised for efficient coupling and reduced absorption losses from the plasmonic nanoparticles. 
The SLR lasers show large-area and single-mode emission with low thresholds down to 94 $\mu$J cm$^{-2}$ pulse$^{-1}$. 
The emission wavelength is tuneable over a broad wavelength range of 70 nm through variation of the nanoparticle array period. 
We theoretically describe the different TE and TM modes of the system both through analytical calculations and numerical simulations.
We obtained an effective cavity radius of 30 $\mu$m by varying the optical pump beam size indicating the cavity coherence length. 
When pumping the device with a beam smaller than the effective cavity, lasing requires higher thresholds leading to a refractive index gradient resulting in mode quantisation.
The semiconductor gain material is not affected by photobleaching, showing stable emission intensity up to 7x10$^8$ excitation pulses. 
The plasmonic-photonic hybrid SLR forms the foundation for further lattice designs and advanced band-edge engineering without relying on thin films of gain material.
Our devices pave the way for on-chip SLR lasing without degradation for applications in optical communication, optical computing and LiDAR.

\subsection*{Methods} \label{sec:methods}

\subsubsection*{Optical measurements}

The SLR lasers are characterised in a micro photoluminescence setup. They are pumped with a 633 nm, 200 fs-pulsed laser at 100 kHz. The pump beam excites the sample from the top and is focused through a 20x objective. The emitted light is collected in reflection through the same objective. The emission spectra are measured in a Princeton Instruments spectrometer. We used a digital micromirror device (DMD) to pump the devices with variable pump diameters for the data in Figure 3.

\subsubsection*{Sample fabrication}

\textbf{InP bonding and \ch{SiO2} spacing layer.}
The active InP layer is integrated on a Si wafer through direct wafer bonding \cite{caimiHeterogeneousIntegrationIII2021}. The layer is grown on a sacrificial, lattice-matched III/V wafer, and then bonded to an Si wafer through an adhesion layer of oxide (which becomes the BOX) and annealing. The sacrificial wafer can then be removed by wet etching. The 50 nm \ch{SiO2} layer separating InP from the gold particles is deposited with Plasma-Enhanced Chemical Vapor Deposition. \\

\textbf{Au nanoparticle fabrication.}
Two layers of resist, polymethyl methacrylate (PMMA) A4 495K and A4 950K, are spun onto the \ch{SiO2} spacing layer consecutively, both at 4000 RPM for 1 minute and each layer baked on a hotplate for 18 minutes at 180$^o$C. A further layer of e-spacer was spun at 2000 RPM for 1 minute with a 30 s bake at 90$^o$C. The resist is exposed to a desired pattern in an electron beam lithography system (EBL, Raith E-line plus). The PMMA is developed in a 3:1 Isopropyl alcohol (IPA) to Methyl isobutyl ketone (MIBK) solution, followed by a 30 rinse in IPA to halt development. Residual PMMA is removed via plasma ashing at 40\% power for 6 s. A 1.5 nm Cr adhesion layer is thermally evaporated followed by a 50 nm layer of Au (Angstrom Engineering Amod). The PMMA and unwanted Cr/Au is removed in the lift-off process, where the sample is left in Acetone for 24 hours. Finally, 60 nm of \ch{SiO2} is sputtered (Angstrom Engineering Amod) onto the Au and spacing layer \ch{SiO2} to encapsulate the Au and create the superstrate. The full set of 36 devices measured is comprised of arrays with periods = 269 $\pm$ 3, 273 $\pm$ 2, 279 $\pm$ 3, 284 $\pm$ 3, 288 $\pm$ 4, 293 $\pm$ 6, 299 $\pm$ 4, 304 $\pm$ 6 and 308 $\pm$ 6; each with nanoparticle diameters = 58 $\pm$ 8, 74 $\pm$ 6, 99 $\pm$ 6 and 126 $\pm$ 5. 

\subsubsection*{Theory}
The lasing modes were analytically calculated using the grating equation, and validated through comparison with numerical simulations using FEM. When assuming normal emission, which is most likely for simple SLRs, and for weak coupling, a simple formula for the mode wavelength can be deduced: $\pm n_{\mathrm{eff}} k_0 = 2\pi/P$, where $n_{\mathrm{eff}}$ denotes the the effective refractive index for excited mode in the slab waveguide, $k_{0}$ denotes the wavenumber of the free space light, and $P$ denotes the period length. The effective refractive index $n_{\mathrm{eff}}$ is numerically calculated using a finite-difference frequency-domain method. To place the theoretically predicted mode wavelengths in a range similar to the experiment, the thickness of the InP layer and the refractive index of the InP were varied. 
Moreover, we performed finite-difference time-domain simulations with ANSYS Lumerical, to obtain the dispersion of the modes of the system and explain additional emission modes.  
More details can be found in the supporting information.

\subsubsection*{Acknowledgments}

AF, JD, HS, and KM acknowledge support from the EU ITN EID project CORAL (GA no. 859841). T.S.M. acknowledges the support of the UK government Department for Science, Innovation and Technology through the UK National Quantum Technologies Programme.
This work is supported by the UK Engineering and Physical Sciences Research Council (EPSRC), through grants: EP/S000755/1, EP/M013812/1 and EP/T027258.
We thank the Cleanroom Operations Team of the Binnig and Rohrer Nanotechnology Center (BRNC) for help with sample fabrication.

\subsubsection*{Author Contributions}

T.S.M. and J.D. fabricated the devices. 
A.F. and T.S.M. took optical measurements. 
X.X. and T.V.R. developed the theory and carried out simulations. 
T.S.M. and A.F. wrote the paper with contributions from all co-authors. H.S., K.M., R.F.O. and R.S. conceived and supervised the project.
T.S.M. and A.F. contributed equally to this work.

\bibliography{SLR_bibliography}
\end{document}